\documentstyle[aps,pre,twocolumn,epsfig]{revtex}
\begin{document}

\twocolumn[\hsize\textwidth\columnwidth\hsize\csname@twocolumnfalse\endcsname
\title{Anisotropic non-perturbative zero modes for passively advected 
magnetic fields}
\author{Alessandra Lanotte$^{1}$ and Andrea Mazzino$^{2,3}$ \\
\small{$^{1}$ CNRS, Observatoire de la C\^ote d'Azur, B.P. 4229,
06304 Nice Cedex 4, France.\\
\small$^2$ INFM-Dipartimento di Fisica, Universit\`a di Genova, 
I--16142  Genova, Italy.}\\
\small$^3$ The Niels Bohr Institute, Blegdamsvej 17, DK-2100 Copenhagen, 
Denmark}
\draft
\date{\today}
\maketitle
\begin{abstract}
A first analytic assessment of the role of anisotropic  corrections to the  
isotropic anomalous scaling exponents is given
for the $d$-dimensional kinematic magneto-hydrodynamics problem in the presence
of a mean magnetic field. The velocity advecting the magnetic field
changes very rapidly in time and scales with a positive exponent $\xi$.
Inertial-range anisotropic contributions to the 
scaling exponents, $\zeta_j$, of second-order magnetic correlations  
are associated to zero modes and have been calculated non-perturbatively.  
For $d=3$, the limit $\xi\mapsto 0$ yields
$\protect{\zeta_j=j-2+ \xi (2j^3 +j^2 -5 j - 4)/[2(4 j^2 - 1)]} $
where $j$ ($j\geq 2$) 
is the order in the Legendre polynomial decomposition applied 
to correlation functions.
Conjectures on the fact that anisotropic
components cannot change the isotropic threshold to the dynamo effect are 
also made.

\end{abstract}
]

Since Kolmogorov \cite{K41} 
formulated his hypothesis, most of theories and models in turbulence have
used as a key ingredient the restored local isotropy of small-scale 
structures, even in the presence of large-scale anisotropy.
Dealing with such an idealized picture implies that 
the anisotropic effects, that almost every large-scale forcing indeed involve,
are totally disregarded.\\
Recently, some considerable effort 
has been done \cite{A98,Sadd94,BoOr96,Arad98,Arad99} 
to shed some light on the statistics of structure functions when taking 
anisotropy  explicitly into account.
When doing this, two major questions emerge:
the first concerns the possibility of an {\it universal} nature of the scaling
exponents of the separated isotropic and anisotropic contributions to 
structure functions;
the second concerns the decay of anisotropic fluctuations, and consequently the validity of the local isotropy hypothesis.
The state of the art on this subject, especially when looking at experimental
data, does not give unique answers.\\
Here, our aim is to analyze, through non-perturbative calculations, 
the effects of anisotropy on anomalous (i.e.~non-dimensional) scaling 
exponents of the magnetic field correlations, within a kinematic 
magneto-hydrodynamics (MHD)
 problem (i.e.~when reaction 
of the magnetic field on the velocity field is neglected).
As the advecting velocity field that we consider 
is $\delta$-correlated in time, analytical 
approach is possible: the main result shown in this Letter is 
that the anomalous scaling exponent, $\zeta_0$, 
associated to the isotropic contribution is dominant with respect to the 
anisotropic ones, $\zeta_j$. 
In addition, the entire set of anisotropic scaling 
exponents, $\zeta_j$, is given, 
showing the existence of a hierarchy related to 
the degree of anisotropy $j$, such that: $\zeta_0 < \zeta_2 <\cdots$.
This result is the signature of the emergence of local isotropy at small 
scales.\\  Such hierarchy relations hold analogous to those
found in Ref.~\cite{A98} where the scalar advection in the presence of 
large-scale anisotropy is studied exploiting the field theoretic
renormalization group (RG).\\
We also remark that the scenario here outlined is compatible with those 
arising from the results shown in Refs.~\cite{BoOr96,Arad99} 
where, respectively, experimental and direct numerical simulation data of 
anisotropic turbulence have been analyzed.
Considering the kinematic MHD problem, the issue of the threshold for the 
appearance on the unbounded growth of the magnetic field is also present. 
We shall report a conjecture according to which we can argue that 
the anisotropic components do not play any effect in this sense.\\
In the presence of a mean component 
 $\bbox{B}^o$ (actually supposed varying 
on very large scales $\sim L$, the largest one in our problem) 
the kinematic MHD equations describing the evolution of the fluctuating 
part, $\bbox{B}$,
of the magnetic field are \cite{Zeldo83}:
\begin{eqnarray}
\label{fp}
&&\partial_t B_\alpha +\bbox{v}\cdot\bbox{\partial}\,
B_\alpha=\bbox{B}\cdot\bbox{\partial}\,v_{\alpha}+\bbox{B}^o\cdot
\bbox{\partial}\,v_{\alpha}+
\kappa\,\partial^2 B_{\alpha}\\
&&\alpha=1,\cdots ,d \nonumber
\end{eqnarray}
where the velocity, $\bbox{v}$,  
is a zero-mean Gaussian random process, homogeneous, isotropic
and white-in-time, $\kappa$ is the magnetic  diffusivity, and
$d$ is the dimension of the space. 
Both $\bbox{v}$ and $\bbox{B}$ are divergence free fields. 
The term  
$\bbox{B}^o\cdot\bbox{\partial}\,v_{\alpha}$ in (\ref{fp}) plays the same
role as an  external forcing driving the system
and being also a  source of 
anisotropy for the magnetic field statistics.\\
The velocity is  self-similar with the 2-point correlation
function:
\begin{equation}
\label{2-point-v}
\langle  v_{\alpha}(\bbox{r},t)v_{\beta}(\bbox{r}',t') \rangle =
\delta(t-t')\,\left[ d^0_{\alpha\beta} -S_{\alpha\beta}(\bbox{r}- \bbox{r}')
\right]   ,
\end{equation}
where $S_{\alpha\beta}(\bbox{r})$ is fixed by isotropy and scaling behavior,
and scales with the exponent $\xi$, in the range $0< \xi< 2$:
\begin{equation}
S_{\alpha\beta}(\bbox{r})=
D r^{\xi}\left [\left (d+\xi-1\right)
\delta_{\alpha\beta} - \xi \frac{r_{\alpha}r_{\beta}}{r^2}
\right]. 
\label{eddydiff}
\end{equation}
The $\delta$-correlation in time of $\bbox{v}$ permits to exploit the Gaussian
integration by parts 
(a comprehensive description of this technique can be found, e.g., 
in Ref.~\cite{F95}) to obtain closed, exact equations for 
$C_{\alpha\beta}(\bbox{r},t)\equiv\langle B_{\alpha} 
(\bbox{x},t) B_{\beta}(\bbox{x+r},t)\rangle$. 
After some manipulations of (\ref{fp}), such equations read:
\begin{eqnarray}
\partial_t\, C_{\alpha\beta}&=&
S_{ij}\,\partial_i\partial_j \,C_{\alpha\beta}-
\left (\partial_j\,S_{i\beta}\right )
\partial_i\,C_{\alpha j}-\nonumber\\
&&\left (\partial_jS_{\alpha i}\right )\left (\partial_i C_{j \beta}\right )+
\left (\partial_i\partial_j S_{\alpha\beta}\right ) 
\left (  C_{ij} + B_i^oB_j^o\right ) + \nonumber \\
&&2\kappa\, \partial^2 C_{\alpha\beta}.
\label{eq-moto}
\end{eqnarray}
A further equation for $C_{\alpha\beta}$ follows from
the solenoidal condition on $\bbox{B}$:
\begin{equation}
 \partial_{\alpha} C_{\alpha\beta} = 0 .
\label{cont}
\end{equation}
For what follows, it is worth emphasizing two properties of $C_{\alpha\beta}$:

\noindent i) because of 
homogeneity, $C_{\alpha\beta}$ is left invariant under the
following set of transformations:
\begin{eqnarray}
\bbox{r}\longmapsto -\bbox{r}\qquad&\mbox{and}&\qquad
\alpha  \longleftrightarrow \beta \label{simme1};
\label{simme2}
\end{eqnarray}
ii) $C_{\alpha\beta}(\bbox{r})=C_{\alpha\beta}(-\bbox{r})$, as it follows
from (\ref{eq-moto}) after the substitution $\bbox{r}\mapsto -\bbox{r}$.

\noindent As shown 
in Ref.~\cite{V96}, in the isotropic case (i.e. $\bbox{B}^o=0$ in our problem)
anomalies
appear already in the scaling exponents of the
second-order magnetic correlations, $C_{\alpha\beta}$, and have been
calculated non-perturbatively by the author.  Anomalous
scaling laws are associated with zero modes of the closed equations
satisfied by the equal-time correlation functions.\\
In Ref.~\cite{AA98}, 
anomalous exponents for higher-order correlations have been calculated
to the order $\xi$  by exploiting the RG technique.

\noindent The ex\-trac\-tion of ani\-so\-tro\-pic contributions 
to the isotropic scaling 
of $C_{\alpha\beta}(\bbox{r})$ found in Ref.~\cite{V96},  
and the investigation of
their effect (if any) on the emergence of the dynamo effect
are the main questions addressed  in the present Letter. 
The main technical difference with respect 
to Ref.~\cite{V96} is that 
the angular structure of zero modes 
has now to be explicitly taken into account.

\noindent In the presence of anisotropy, the most general
expression for the 2-point magnetic correlations,
$C_{\alpha\beta}(\bbox{r})$, in the stationary state involves five
(two in the isotropic case) functions depending on both $r\equiv
|\bbox{x}-\bbox{x}'|$ and $z\equiv
\cos\theta=\hat{\bbox{B}}^o\cdot\bbox{r}/r$, where $\hat{\bbox{B}}^o$
is the unit vector corresponding to the direction selected by the mean
magnetic field.
Remark that the space is anisotropic but still homogeneous,
so there is no explicit dependence on the points $ \bbox{x},\bbox{x}'$,
but only on their distance.
 Namely,
\begin{eqnarray}
C_{\alpha\beta}(\bbox{r})
& =& {\cal F}_1(r,z)\;\delta_{\alpha\beta} + 
{\cal F}_2(r,z)\frac{r_{\alpha}r_{\beta}}{r^2} +\nonumber\\
&& {\cal F}_3(r,z)\frac{\hat{B}^o_{\alpha} r_{\beta}}{r}+ 
{\cal F}_4(r,z)\frac{\hat{B}^o_{\beta} r_{\alpha}}{r}+\nonumber\\
&& {\cal F}_5(r,z)  \hat{B}^o_{\alpha}\;\hat{B}^o_{\beta}.
\label{general}
\end{eqnarray}
 From the properties i) and ii) of $C_{\alpha\beta}(\bbox{r})$ one immediately obtains
the following relations for the ${\cal F}$'s:
\begin{eqnarray}
{\cal F}_i(r,z)&=&{\cal F}_i(r,-z)\qquad i=1,2,5 \label{rela}\\
{\cal F}_3(r,z)&=&-{\cal F}_3(r,-z)\label{reld}\\
{\cal F}_3(r,z)&=&{\cal F}_4(r,z)\label{rele}.
\end{eqnarray}
Substituting the expression (\ref{general}) into (\ref{eq-moto}) and using the 
chain rules, we obtain, after lengthy but straightforward algebra, the following four equations
(corresponding to the projections over $\delta_{\alpha\beta}$, 
$r_{\alpha}r_{\beta}/r^2$, $\hat{B}^o_{\alpha} r_{\beta}/r$
and $\hat{B}^o_{\alpha}\;\hat{B}^o_{\beta}$):
\begin{eqnarray}
%
%
&[&a_1r^2\partial^2_r+b_1 r\partial_r + 
c_1 (1-z^2)\partial^2_z+
d_1 z\partial_z +e_1] {\cal F}_1+\nonumber\\
&[&f_1 r\partial_r+g_1 z\partial_z + j_1]\,{\cal F}_2 +
[k_1 z\, r\partial_r +l_1z^2\partial_z + m_1z]{\cal F}_3+\nonumber\\
&[&o_1 +p_1z^2]\,{\cal F}_5 = (q_1+r_1z^2)\,B^{o\;2}\label{eq1}\\
&&\vspace{-10mm}\nonumber\\
%
%
&a&_2 {\cal F}_1+\nonumber\\
&[&b_2r^2\partial^2_r+c_2 r\partial_r + 
d_2 (1-z^2)\partial^2_z+
e_2 z\partial_z +f_2]\, {\cal F}_2+\nonumber\\
&g&_2 z\,{\cal F}_3+ 
\left[k_2 +l_2z^2 \right]\,{\cal F}_5 = (m_2+n_1z^2)\,B^{o\;2}\label{eq2} \\
&&\vspace{-10mm}\nonumber\\
&a&_3 \partial_z  {\cal F}_1+b_3\partial_z  {\cal F}_2+\nonumber\\
&[&c_3r^2\partial^2_r + d_3 r\partial_r + e_3 (1-z^2)\partial^2_z+
f_3 z\partial_z + g_3 ]\,{\cal F}_3+\nonumber\\
&[&j_3 z\,r\partial_r+(k_3+l_3z^2)\partial_z+m_3 z]\,{\cal F}_5
= n_3\,B^{o\;2}\, z\label{eq3}\\
&&\vspace{-10mm}\nonumber\\
%
%
&a&_4\partial_z  {\cal F}_3+ \nonumber\\
&[&b_4 r^2\partial^2_r + c_4 r\partial_r + d_4 (1-z^2)\partial^2_z+
e_4 z\partial_z+f_4]\,{\cal F}_5=\nonumber\\
&g&_4\, B^{o\;2},\label{eq4}
\end{eqnarray}
where the coefficients $a_i, b_i, \cdots r_i$ are cumbersome functions of 
$\xi$ and $d$
and will not be here reported for the sake of brevity. Without loss 
of generality, we have fixed $D=1$ in (\ref{eddydiff}), and we have neglected all terms involving 
the magnetic diffusivity $\kappa$, our attention being indeed focused in the 
inertial range of scales, i.e.~$\eta \ll r\ll L$ 
where $\eta=\kappa^{1/\xi}$ is the dissipative scale for the problem. \\
With the substitution of the expression (\ref{general}), 
the solenoidal condition  (\ref{cont}) splits into
the following couple of equations:
\begin{eqnarray}
&[&r\partial_r +(d-1)]\, {\cal F}_1 +[r\partial_r 
-z\partial_z]\, {\cal F}_2+\nonumber\\
&[&z\,r\partial_r +\partial_z -
z^2\partial_z -z]\, {\cal F}_3=0\label{eqcont1}\\
&&\vspace{-10mm}\nonumber\\
&\partial&_z {\cal F}_2+ [r\partial_r + d]\,{\cal F}_3+
[z\,r\partial_r + (1-z^2)\partial_z]\,{\cal F}_5 = 0 \label{eqcont2}
\end{eqnarray}
associated to the projections over $r_{\beta}/r$ and $\hat{B}^o_{\beta}$, 
respectively.\\
 From the relation (\ref{rele}) and Eqs.~(\ref{eqcont1}) and (\ref{eqcont2}) 
it then follows 
that only two functions, ${\cal F}'s$, in (\ref{general}) are independent.\\
According to the old idea of Kolmogorov, in cascade-like mechanisms
 of transfer of energy towards small scales, 
anisotropy present at the
integral scale should eventually decay during the (chaotic) transfer. One could thus argue 
that, at least at small scales,  
anisotropic corrections to the isotropic contribution
would be smaller and smaller as the order of anisotropic contributions
increases. For Navier--Stokes turbulence in channel flow,
such a picture has recently been confirmed 
by Arad {\it et al} in Ref.~\cite{Arad99}.\\
As we shall see, the above physical hint can be easily
exploited if one decomposes functions ${\cal F}'s$ 
on the Legendre polynomial basis. Accordingly, we have:
\begin{eqnarray}
&{\cal F}&_i(r,z)=\sum_{l=0}^{\infty} f_{2l}^{(i)}(r)\, 
P_{2l}(z)\qquad i=1,2,5 \label{f125}\\
&{\cal F}&_3(r,z)=\sum_{l=0}^{\infty} f_{2l+1}^{(3)}(r)\, P_{2l+1}(z), 
\label{f3}
\end{eqnarray}
where the separation of even and odd orders in (\ref{f125}) and (\ref{f3})  arises as a consequence of the symmetries expressed by the relations (\ref{rela}) and (\ref{reld}), 
respectively. As larger $l$'s correspond to higher order anisotropic
contributions, we thus expect that, when scaling behavior sets in 
(i.e.~for $\eta \ll r\ll L$), we shall have:
\begin{equation}
f_l^{(i)}(r) \propto 
r^{\zeta_l^{(i)}}\qquad\mbox{with}\qquad\zeta_0^{(i)}<\zeta_1^{(i)}<\cdots .
\label{decay}
\end{equation}
We would like to obtain equations for the $f^{(i)}_l(r)$, to be then solved for the
$\zeta_l^{(i)}$: to do this, in the Eqs.~(\ref{eq1})-(\ref{eqcont2}) we have to express  quantities like $z^j\partial_z^m$
($j,m=0,1,2$,  with $j\neq1$ and  $m\neq 2$) in terms of Legendre polynomials. 
Recalling to this purpose
well-known relations involving these latter
(see, e.g., Ref.~\cite{Grad65}), we obtain general expressions as e.g.:
\begin{equation}
\partial_z {\cal F}_i(r,z)=\sum_{l=0}^{\infty}P_l(z)\,\left[(2l+1)
\sum_{q=0}^{\infty}f_{2q+l+1}^{(i)}(r)\right],
\label{esempio}
\end{equation}
from which we notice that an arbitrary $l$-order
anisotropic contribution is
coupled to all larger orders. The resulting equations 
arising from (\ref{eq1})-(\ref{eqcont2}) are thus not closed.\\
Closed equations for the $f$'s can actually be obtained by 
exploiting (\ref{decay}), i.e.~by using the hypothesis of a hierarchy 
in the self-similar behavior of the $f$'s. Accordingly, in 
Eqs.~(\ref{eq1})--(\ref{eqcont2}) at a given order $j$, for each function 
$f_l^{(i)}$  we need to retain only its 
lower order contributions with $l \leq j$. It is worth noticing that we can 
control the validity of this (physical) assumption in a self-consistent 
way, at the end of our calculation. \\
As a result, one obtains the following
(infinite) set of closed differential equations, valid for $j$ even:
\begin{eqnarray}
%
%
&a&_1 r^2 f_j^{''(1)} + b_1 r f_j^{'(1)} + c_1  f_j^{(1)} +  
d_1 r f_j^{'(2)} + e_1  f_j^{(2)} + \nonumber\\
 &g&_1 r f_{j-1}^{'(3)} + 
j_1  f_{j-1}^{(3)} + k_1  f_{j-2}^{(5)} = 
B^{o\;2}\,l_1\,\delta_{j2} \label{neweq1}\\
&&\vspace{-10mm}\nonumber\\
%
%
&a&_2 f_j^{(1)} + 
b_2 r^2 f_j^{''(2)} + c_2 r f_j^{'(2)} + d_2  f_j^{(2)} +
e_2 f_{j-1}^{(3)} + \nonumber\\
&g&_2 f_{j-2}^{(5)} = B^{o\;2}\,j_2\,\delta_{j2} \label{neweq2}\\  
&&\vspace{-10mm}\nonumber\\
%
%
&a&_3 f_j^{(1)} + b_3 f_j^{(2)} +
c_3 r^2 f_{j-1}^{''(3)} + d_3 r f_{j-1}^{'(3)} + e_3  f_{j-1}^{(3)} + 
\nonumber\\
&g&_3 r f_{j-2}^{'(5)} + j_3 f_{j-2}^{(5)} = B^{o\;2}\,k_1\,\delta_{j2}
\label{neweq3}\\
&&\vspace{-10mm}\nonumber\\
%
%
&a&_4 f_{j-1}^{(3)} +
b_4 r^2 f_{j-2}^{''(5)} + c_4 r f_{j-2}^{'(5)} + d_4  f_{j-2}^{(5)} =
B^{o\;2}\,e_4\,\delta_{j2}\label{neweq4}\\  
&&\vspace{-10mm}\nonumber\\
&r& f_{j}^{'(1)}+ (d-1) f_{j}^{(1)} +
r f_{j}^{'(2)} - j f_{j}^{(2)} +\nonumber\\
\lefteqn{\frac{j}{2j-1} r f_{j-1}^{'(3)} - \frac{j^2}{2j-1} f_{j-1}^{(3)} = 0}
\label{neweqcont1}\\
&&\vspace{-10mm}\nonumber\\   
&(&2j-1)f_{j}^{(2)} + r f_{j-1}^{'(3)} +d\, f_{j-1}^{(3)} 
+ r \frac{j-1}{2j-3}f_{j-2}^{'(5)}-\nonumber\\
\lefteqn{\frac{(j-2)(j-1)}{2j-3}f_{j-2}^{(5)}=0,}\label{neweqcont2}
\end{eqnarray}
where coefficients $a_i$, $b_i$, $\cdots$ (different from those defined
in (\ref{eq1})-(\ref{eq4})) depend only on $\xi$ and $d$. For $j=0$, the
coefficients relative to $f_{j-1}^{(3)}$ and $ f_{j-2}^{(5)}$ 
(and their derivatives) are zero and the resulting equations for 
$f_{0}^{(1)}$ and $f_{0}^{(2)}$ are exactly  as in Ref.~\cite{V96}.
%
%
The structure of the above equations fixes the relation between the scaling 
exponents relative to different $f$'s. Indeed, when searching for
power law  solutions $f^{(i)}_j(r)\propto r^{\zeta_j^{(i)}}$, 
in order to obtain  balanced equations the `oblique' 
relations must hold:
\begin{equation}
\zeta_j\equiv \zeta_j^{(1)}=\zeta_j^{(2)}=\zeta_{j-1}^{(3)}=\zeta_{j-2}^{(5)}.
\label{obliqua}
\end{equation}
\noindent We are now ready to show that nontrivial scaling behaviors for the $f$'s 
take place due to zero modes, i.e.~the solutions of the homogeneous problem
associated to Eqs.~(\ref{neweq1})-(\ref{neweqcont2}). 
To that purpose, let us consider such 
differential problem with no forcing (i.e.~$B^{o}=0$).
If, when looking for power law solutions, we exploit (\ref{obliqua}) 
and the fact that only two functions of the $f's$ are independent, our 
differential problem is mapped into an algebraic one. In this latter 
the emergence of 
zero modes reduces to imposing the existence of non-zero solutions
of a $2\times 2$ homogeneous linear system. The calculation is lengthy
but  straightforward and the following expressions for the 
zero-mode exponents are obtained\cite{grazie}:
\begin{eqnarray}
\zeta_j^{\pm} &=&  - \frac {1}{2(d-1)} \left\{2\,\xi + d^{2} - d - 
\left[ - 2\,d^{3}\,\xi  - 2\,d^{2}\,\xi^{2} -\right .\right .\nonumber\\
&6&\hspace{-1mm}d^{3} + 4\, \xi^{2}\,d + 8 
+ 10\,d\,
\xi  + 20\,d\,j - 20\,d - 8\,\xi  - \nonumber\\
&8&\hspace{-1mm} j + 4\,d^{2}\,j^{
2} + 2\,\xi^{2}- 4\,\xi\,j^{2} + 17\,d^{2} - 8\,d\,j^{2} + \nonumber\\
&8&\hspace{-1mm} \xi \,j + 4\,d^{3}\,j + 4\,d^{2}\,j\,\xi 
 + 4\,d\,j^{2}\,\xi  + 4\,j^{2} - 16\,d^{2}\,j -\nonumber\\
&1&\hspace{-2mm}2\,d\,\xi\,j + d^{4} \pm 
\left . \left . 2\sqrt{K} \left( d-1\right) \left( 2-\xi\right) 
\right]^{1/2}\right\} 
\label{amore}
\end{eqnarray}
where 
\begin{eqnarray}
 K &=& (d - 1)\,(d^{3} + 4\,d^{2}\,j - 5\,d^{2} + 2\,d^{
2}\,\xi + \xi^{2}\,d + 4\,d\,\xi 
\,j - \nonumber\\ 
&6&\hspace{-0.6mm}d\,\xi  + 8\,d - 12\,d\,j + 4\,d\,j^{2} - \xi^{2} + 4\,\xi 
 + 8\,j - 8\,\xi \,j - \nonumber\\
&4&\hspace{-1.4mm} - 4\,j^{2} + 4\,\xi \,j^{2}).\nonumber
\end{eqnarray}
Some remarks are in order. First, $\zeta_0^{+}$ coincides with the isotropic 
solution obtained by Vergassola in Ref.~\cite{V96}, the admissibility of which has been
proved by the author. Second, $\zeta_0^{-}$ diverges as $r^{-2}$ at the 
dissipative scale $\eta$. Exponent  $\zeta_0^{-}$ is thus not admissible.
Third, $\zeta_j^{+} > \zeta_j^{-}$ for all $j$. This means that, in the 
inertial range of scales (i.e.~$r/L\ll 1$) the leading zero-mode solutions 
for $j\geq2$ are associated to $\zeta_j^{-}$. We can thus define the leading set
of zero modes $\zeta_j$ as: $\zeta_0 \equiv \zeta_0^{+}$;
$\zeta_j \equiv \zeta_j^{-}$ for $j\geq 2$.

\noindent In particular, for $j=2$, 
the asymptotic limits $\xi\ll 1$ and $d\gg 1$ are, respectively:
\begin{equation}
\zeta_2=\frac{2\xi}{(d-1)(d+2)}+ O(\xi^2);~\zeta_2 = \frac{2\xi}{d^2}+O(1/d^3).
\end{equation}
\noindent Let us briefly discuss the infrared (IR) behavior of $\zeta_j$.
In the absence of forcing terms, there is no way to 
satisfy the IR boundary condition 
(i.e.~$C_{\alpha\beta}(\bbox{r})=0$ for $r\mapsto \infty$): 
$\zeta_j$
indeed diverges for $r\mapsto \infty$. As a consequence, 
zero modes for $j>2$ are not globally acceptable.
For  $j=2$ the situation changes completely: in this case 
Eqs.~(\ref{neweq1})-(\ref{neweq4}) are forced and, as in Ref.~\cite{V96},
IR boundary 
conditions can be satisfied by matching at the large scale $L$ 
zero-mode solutions with those of the inhomogeneous problem.
 From the above considerations, it also follows that
zero modes associated to $\zeta_j$ become acceptable for all orders $j$
when a fully anisotropic forcing term (i.e.~projecting on all Legendre 
polynomials) is added in the right hand side of (\ref{fp}).

\noindent Finally, autoconsistency of our solution for $\zeta_j$, 
that is the validity of the hierarchy in 
(\ref{decay}), can be immediately checked from Fig.~1
where the behaviors of few $\zeta_j$ as functions of $\xi$ are shown in the 
three dimensional case for $j=0,2,4$ and $6$ (from below to above).
It is easy to verify that the increasing of scaling exponents with $j$
actually holds for all values of $j$ and $d$.
\begin{figure}
\centerline{\psfig{file=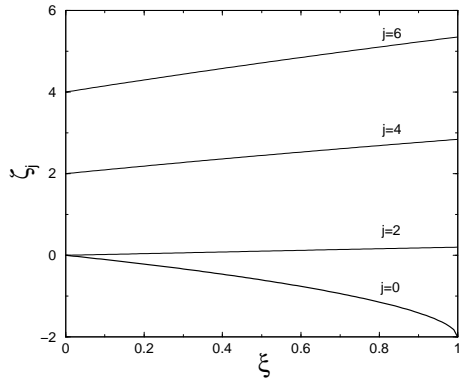,width=6cm}}
\caption{Behavior of $\zeta_j$ 
{\it vs} $\xi$ for $d=3$ and (from below to above) $j=0$ (heavy line)
$j=2,4$ and $6$ (thin lines).}
\end{figure}
The expression for $\zeta_j$ allows us to make some conjectures on the
role played by anisotropic effects on the emergence of the
dynamo effect.
It is known \cite{V96,KA68} that in the isotropic case
an unbounded growth of the magnetic 
field (dynamo effect) arises for $\xi>1$. The question addressed here
is whether anisotropic component can contribute 
to destabilize the system,  
shifting toward smaller values of $\xi$ the threshold for the dynamo.
We note that,
in the isotropic case, dynamo arises when the exponent related to the 
admissible zero mode becomes complex. 
In this case, zero-mode solutions have sinusoidal components, a fact that
makes possible their matching with the appropriate boundary conditions also 
in the absence of forcing (i.e.~the system is self-maintained).
This happens for 
$\xi>1$, $\xi=1$ being the threshold.  Taking such condition 
as a criterion to select the emergence
of an unbounded growth, we can conclude that there is no effect played
by the anisotropic components. Indeed, it is easily verified 
from (\ref{amore}) that, for all $d$'s,
$\zeta_j$ is real for  $\xi\in [0,1]$.
\vspace{2mm}
In conclusion, we have presented a system where the extraction of 
anisotropic contributions to the anomalous scaling of the 
equal-time magnetic correlation functions
can be performed in a nonperturbative way. We have
calculated the entire set of universal anomalous exponents, $\zeta_j$,
and we have given
an analytic assessment of the dominance of the fundamental
exponent associated to the isotropic shell. More generally, the hierarchy
$\zeta_0<\zeta_2< \cdots \zeta_j< \cdots$ has been proved.
The picture here drawn is in agreement with recent findings 
by Antonov in Ref.~\cite{A98}, where the passive scalar problem is
studied, and by Arad {\it et al} in Ref.~\cite{Arad99} 
for Navier--Stokes  turbulence.

\bigskip

It is a pleasure to thank A.~Gruzinov and M.~Vergassola
for their stimulating suggestions on the subject matter.
Useful suggestions from N.~Antonov, L.~Biferale, 
A.~Celani, R.~Festa and J.-L.~Gilson are also acknowledged.
AM gratefully acknowledges the EU for the research contract 
n.~FMRX-CT-98-0175 `Intermittency in Turbulent Systems', 
which partially supported  the present work.
AL is supported by the EU contract n.~ERB-FMBI-CT96-0974.

\end{document}